\documentclass[aps,showpacs,amsmath,amssymb,showkeys]{revtex4}
\usepackage{graphics}
\usepackage{bm}
\usepackage{graphicx}% Include figure files
\usepackage{bm}% bold math
\usepackage{color}

\begin{document}

\newcommand{\be}{\begin{equation}}
\newcommand{\ee}{\end{equation}}
\newcommand{\ben}{\begin{eqnarray}}
\newcommand{\een}{\end{eqnarray}}
\newcommand{\nn}{\nonumber \\}
\newcommand{\ii}{\'{\i}}
\newcommand{\pp}{\prime}
\newcommand{\tr}{{\mathrm{Tr}}}
\newcommand{\nd}{{\noindent}}
\newcommand{\grad}{\hspace{-2mm}$\phantom{a}^{\circ}$}

\title{The thermal statistics of quasi-probabilities' analogs in phase space}

\author{F.~Pennini$^{1,2}$,  A.~Plastino$^{3}$, M.C. Rocca$^{3}$}

% Affiliations / Addresses
\address{$^{1}$Departamento de F\'{\i}sica, Facultad de Ciencias Exactas y Naturales,
Universidad Nacional de La Pampa, Av. Peru 151, 6300, Santa Rosa,
La Pampa, Argentina \\$^{2}$Departamento de F\'{\i}sica, Universidad Cat\'olica del Norte, Av.~Angamos~0610, Antofagasta, Chile\\
$^{3}$Instituto de F\'{\i}sica La Plata--CCT-CONICET, Universidad Nacional de La Plata, C.C.~727, 1900,
La Plata, Argentina }

\date{\today}

\date{Received: date / Revised version: date}
\begin{abstract}
We focus attention upon the thermal statistics of the classical
analogs of  quasi-probabilities's (QP) in phase space for the
important case of quadratic Hamiltonians. We consider  the three
more important OPs: 1) Wigner's, $P$-, and Husimi's. We show that,
for all of them, the ensuing semiclassical entropy is a function
{\it only} of the fluctuation product $\Delta x \Delta p$. We
ascertain that {\it the semi-classical analog of the
$P$-distribution} seems to become un-physical at very low
temperatures.  The behavior of several other information
quantifiers reconfirms such an assertion in manifold ways.  We
also examine the behavior of the statistical complexity and of
thermal quantities like the specific heat.
\end{abstract} % end of abstract


% {03.65.Sq}{Semiclassical theories and applications}   \and
%      {05.30.-d}{Quantum statistical mechanics} \and {03.67.-a}{Quantum information}
\maketitle

 %%%%%%%%%%%%%%%%%%%%%%%%%%%%%%%%%%%%%%%%%%%%%%%%%%%%%%%%%%%%%%%
\section{Introduction}
 %%%%%%%%%%%%%%%%%%%%%%%%%%%%%%%%%%%%%%%%%%%%%%%%%%%%%%%%%%%%%%%

A quasi-probability distribution is a mathematical construction
that resembles  a probability distribution but does not
necessarily fulfill some of the Kolmogorov's axioms for
probabilities~\cite{plato}.
 Quasi-probabilities exhibit general features of
ordinary probabilities. Most importantly, they yield expectation
values with respect to the weights of the distribution. However,
they  disobey  the third probability postulate \cite{plato}, in the
sense that regions integrated under them do not represent
probabilities of mutually exclusive states. Some
quasi-probability distributions exhibit zones of negative
probability density. This kind of  distributions often arise in
the study of quantum mechanics when discussed in a phase space
representation, of frequent  use in quantum optics, time-frequency
analysis, etc.

One usually considers a density operator $\hat{\rho}$, defined
with respect to a complete orthonormal basis and  shows that it
can always be written in a diagonal manner, provided that  an
overcomplete basis is ta hand \cite{sudar}. This is the case of
coherent states $|\alpha\rangle$ \cite{glauber}, for
which~\cite{sudar}

\be  \label{UNO} \hat{\rho}= \int \frac{d^2\alpha}{\pi}\,
P(\alpha,\alpha^*) \, \vert \alpha \rangle\langle\alpha\vert. \ee
 We have  $\mathrm{d}^2
\alpha/\pi=\mathrm{d}x\mathrm{d}p/2\pi\hbar$, with $x$ and $p$
phase space variables.  Coherent states, right eigenstates of the
annihilation operator $\hat{a}$, serve as the overcomplete basis
in such a build-up~\cite{sudar,glauber}.

There exists a family of different representations, each connected
to a different ordering of the creation and destruction operators
$\hat{a}$ and $\hat{a}^\dagger$. Historically, the first of these
is the Wigner quasi-probability distribution $W$ \cite{wigner},
related to symmetric operator ordering. In quantum optics the
particle number operator is naturally expressed in normal order
and, in the pertinent scenario, the associated representation of
the phase space distribution is the Glauber--Sudarshan $P$
one~\cite{glauber}. In addition to $W$ and $P$, one may find many
other quasi-probability distributions emerging in alternative
representations of the phase space distribution \cite{smooth}. A
quite  popular representation is the Husimi $Q$
one~\cite{husimi,mizrahi1,mizrahi2,mizrahi3}, used when operators
are in anti-normal order. We emphasize that we work here with
classical analogs of $W,\, P$, and $Q$. As stated, we will
specialize things to the three $f-$functions associated to a
Harmonic Oscillator~(HO) of angular frequency $\omega$. In such a
scenario the three (classical analog)  functions --that we call
for convenience $f_P$, $f_Q$, and $f_W$-- are just Gaussians. The
pertinent treatment becomes wholly analytical.

\subsection{Our goal}

In this paper we wish to apply {\it semiclassical information
theory tools} associated to  these analog $P$, $Q$, and $W$
representations (for quadratic Hamiltonians) {\it in order to
describe the associated thermal semiclassical features}. The idea
is to gain physical insight from the application of different
information-quantifiers to classical analogs of quasi-probability
distributions. It will be seen that useful insights are in this
way gained. WE will discover that out of  the three   functions, only  $f_Q$ and $f_W$ are sensible analogs, while  $f_P$ exhibits problems if the temperature is low enough.

We insist. In his paper we will regard quasi-probabilities as
semi-classical distributions in phase space --analogs of the
quantum quasi-probabilistic distributions--, and try to ascertain
what physical features are they able to describe at such
semi-classical level. One has~\cite{librazo,bookQOptics}:

 \ben
f_P=\gamma_P\,
e^{-\gamma_P|\alpha|^2},\,\,\,\gamma_P&=&e^{\beta\hbar\omega}-1\,\,
\,\,  (P-\textrm{function}),\\\cr f_Q=\gamma_Q\,
e^{-\gamma_Q|\alpha|^2},\,\,\,\gamma_Q&=&1-e^{-\beta\hbar\omega}\,\,
\,\,(Q-\textrm{function}),\\\cr f_W=\gamma_W\,
e^{-\gamma_W|\alpha|^2},\,\,\,\gamma_W&=&2\tanh(\beta\hbar\omega/2)\,\,
\,\,  (W-\textrm{function}),\label{3gamas} \een
 with $\beta=1/k_B T$,  $k_B$ the Boltzmann constant, and $T$ the temperature.
 They will be used  as
 semiclassical statistical weight functions.  Since ours is not a quantum approach, the ordering
 of the HO-creation and destruction operators $\hat a$ and  $\hat a^{\dagger}$ plays no role whatsoever.

\subsection{Historic considerations and organization}
The thermodynamics properties associated to coherent states have
been the subject of much interest. See, for instance, Refs.
\cite{1} and \cite{lieb}).   Notice that the HO is a really
relevant system that yields useful insights of wide impact.
Indeed, the HO constitutes much more than a mere  example. It is
of special relevance for  bosonic or fermionic atoms contained in
magnetic traps~\cite{a6,a7,a8}, as well as for system that exhibit
an equidistant level spacing in the vicinity of the ground state,
like nuclei or Luttinger liquids.

For thermal states, the   gaussian HO-quantum phase spaces
distributions  are known in the literature for applications in
quantum optics.

This paper is organized as follows: section \ref{quantifiers}
refers to  different information quantifiers  in a phase space
representation for Gaussian distributions. In Section \label{analogfanor} we calculate the classical analog-Fano factor. Features of the
fluctuations are analyzed in Section \ref{uncert}. Also, we
discuss the notion of  linear entropy.  Finally, some
conclusions are drawn in Section~\ref{conclu}.

\section{Semi-classical information quantifiers}
\label{quantifiers} Consider a general normalized gaussian
distribution in phase space

\be f(\alpha)=\gamma\,e^{-\gamma |\alpha|^2},\label{gaus} \ee
whose normalized  variance is $1/\gamma$ and $\gamma$ taking
values $\gamma_P$, $\gamma_Q$ and $\gamma_W$. We discuss next, in
these terms, some important information theory quantifiers.

\subsection{ Shift-invariant Fisher's information measure}

\begin{figure}[h]
\begin{center}
\includegraphics[scale=0.6,angle=0]{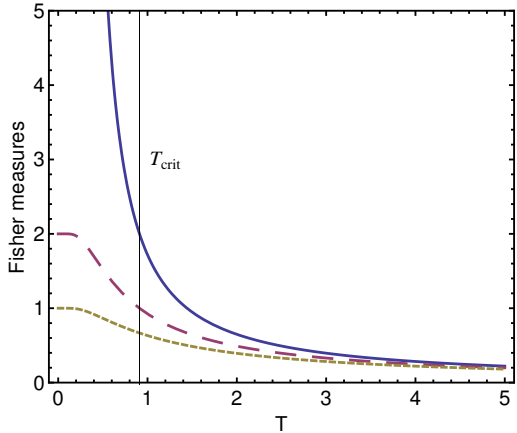}
\vspace{-0.2cm} \caption{Fisher measure versus temperature $T$,
expressed in $(\hbar \omega/k_B)-$units. The solid line is the case $P$,
the dashed one the Wigner one, and the dotted curve is assigned
to the Husimi instance. The vertical line represents the critical
 temperature $T_{crit}$.}\label{fig_fisher}
\end{center}
\end{figure}

The information
quantifier  Fisher's information measure, specialized for families
of  shift-invariant distributions, that do not change shape under
translations, is \cite{bernie,hall} \be I=\,\int
dx\,f(x)\,\left(\frac{\partial \ln f(x)}{\partial x}\right)^2, \ee
and, in phase space, adopts the appearance \cite{Entropy2014}

\be I=\frac14\,\int
\frac{\mathrm{d}^2\alpha}{\pi}\,f(\alpha)\,\left(\frac{\partial
\ln f(\alpha)}{\partial |\alpha|}\right)^2, \ee
such that considering $f(\alpha)$ given by Eq. (\ref{gaus}) we get $I=\gamma$,
 whose
specific values are $\gamma_P$, $\gamma_Q$, $\gamma_W$ for the
three functions $f_P$, $f_Q$, and $f_W$. The behavior of these
quantities are displayed  in Fig. \ref{fig_fisher}. The solid line
is the case P, the dashed one the Wigner one, and the dotted curve
is assigned to the Husimi case. Now, it is known that in the present
scenario the maximum attainable value for $I$ equals 2~\cite{Entropy2014}. The $P$-result violates this restriction at low
temperatures, more precisely at

\be T< T_{crit} = (\hbar\omega/k_B)/\ln{3}\approx 0.91023 \hbar\omega/k_B, \label{crit} \ee with
$T$ being expressed in $(\hbar \omega/k_B)-$units.

\subsection{Logarithmic entropy $S$}

The logarithmic Boltzmann's  information measure for the the
probability distribution  (\ref{gaus}) is

\be S=-\int\,\frac{\mathrm{d}^2 \alpha}{\pi}\,f(\alpha)\,\ln
f(\alpha)=1-\ln \gamma,\label{Shannon} \ee so that it acquires the
particular values

\begin{figure}[h!]
\begin{center}
\includegraphics[scale=0.6,angle=0]{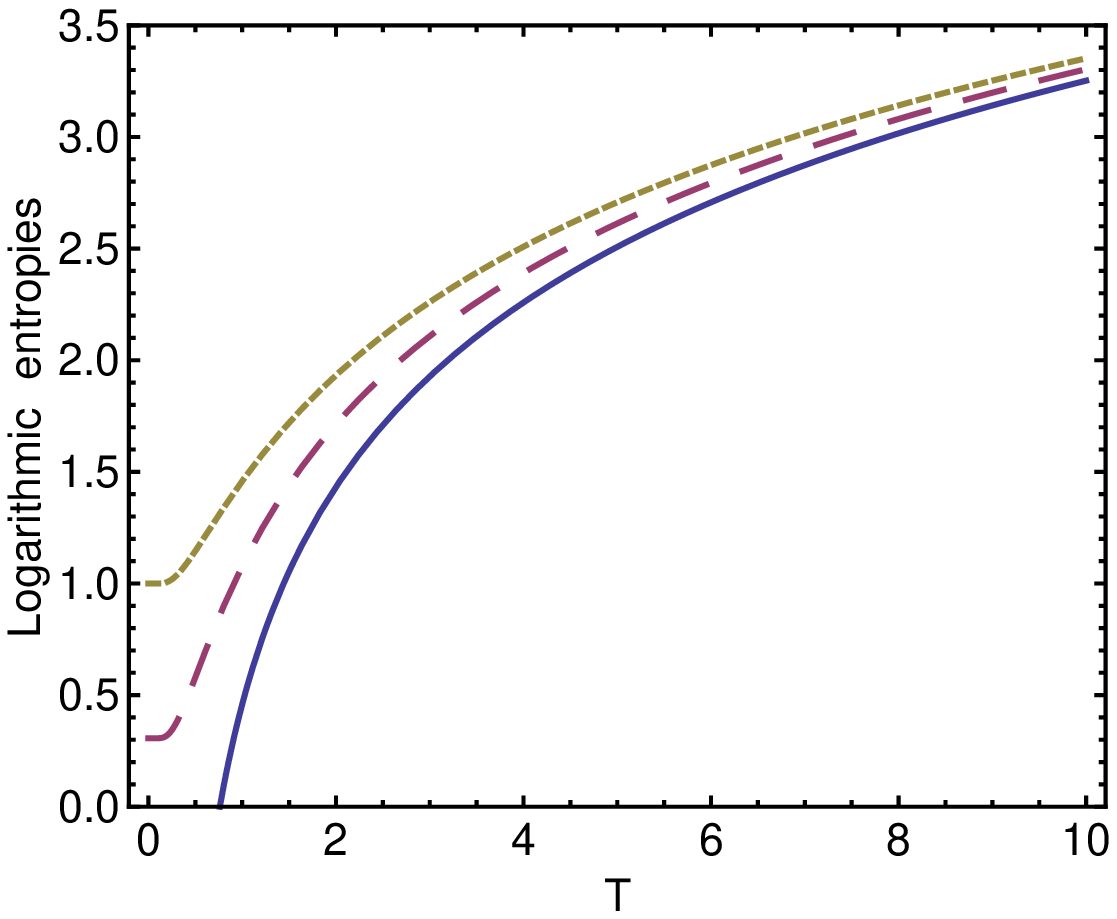} \includegraphics[scale=0.6,angle=0]{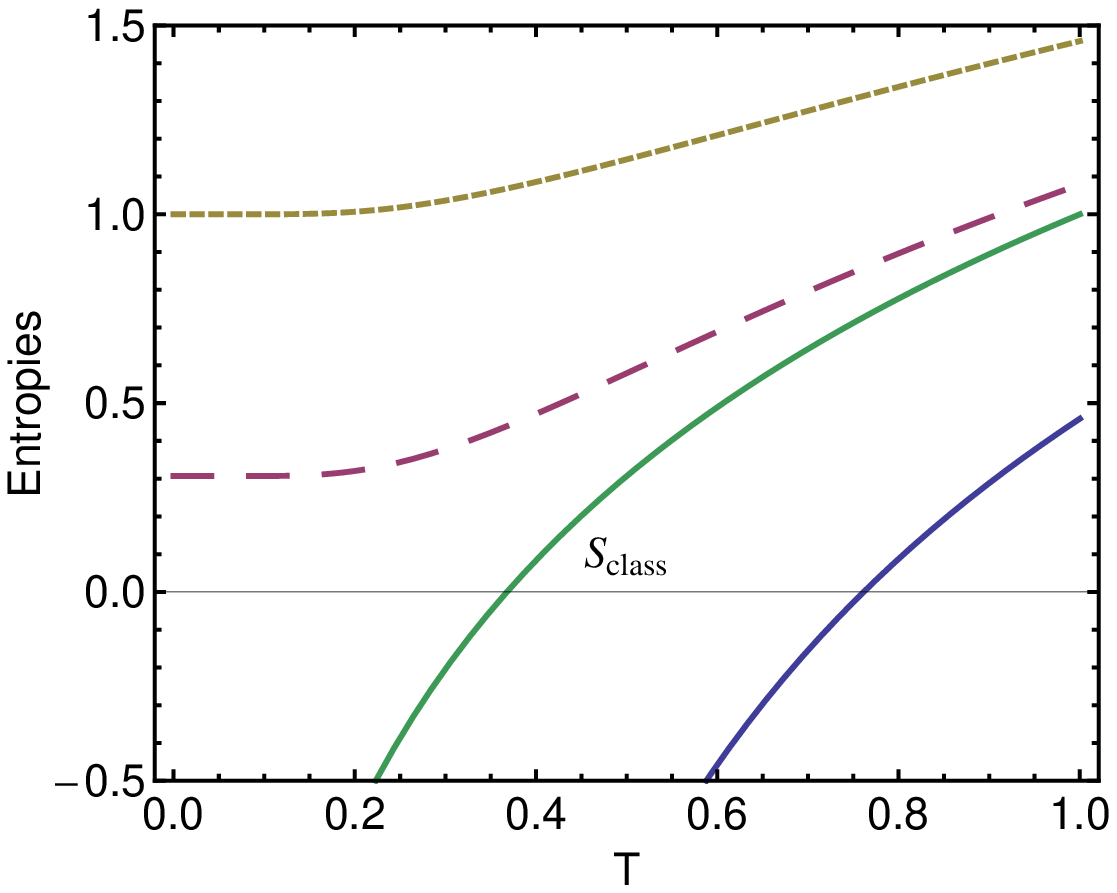}
\vspace{-0.2cm} \caption{Left: logarithmic entropies $S_P$, $S_Q$,
and $S_W$, as a function of the temperature $T$ in $(\hbar
\omega/k_B)-$units. The solid line is the case $P$, the dashed one
the Wigner one, and the dotted curve is assigned to the Husimi
instance. Right:  zoom of the logarithmic entropies as a function
of temperature $T$ in $(\hbar \omega/k_B)-$units. Negative values
of $S_P$ occur below $T=\hbar\omega/(k_B \ln(1+e)$, the critical
temperature~$<T_{crit}$. Remaining details are similar to those of
left figure. We have added the classical entropy of the harmonic
oscillator
$S_{class}=1-\ln(\beta\hbar\omega)$.}\label{fig_entropic}
\end{center}
\end{figure}

\ben S_P&=&1-\ln\left(e^{\beta\hbar\omega}-1\right),\\\cr
S_Q&=&1-\ln\left(1-e^{-\beta\hbar\omega}\right), \\\cr
S_W&=&1-\ln\left(2 \tanh(\beta\hbar\omega/2)\right), \een for,
respectively, the distributions $f_P$, $f_Q$, and $f_W$. These
entropies are plotted in Fig. \ref{fig_entropic}.  Notice that $S_P<0$ for
$T<\hbar\omega/(k_B \ln(1+e)) \approx 0.76 (\hbar\omega/k_B)   <
T_{crit}$. Negative classical entropies  are well-known. One can
cite, as an example, Ref.~\cite{lieb1}.

Table
\ref{tablec} lists a set of critical temperatures $ T_{crit}$
 for typical electromagnetic (EM) waves.

\begin{table}[h]
\begin{tabular}{|l|l|l|}
\hline
   &\textrm{frequency} ($\nu$)&\textrm{Critical temperatures} ($^{\circ}K$)\\
   \hline
  \textrm{Extremely low frequency ELF} &$ 3-30 Hz$& $1.4397\, 10^{-10}-1.4397\, 10^{-9}$ \\
  \textrm{Super low frequency SLF} & $30-300 Hz$ & $1.4397\, 10^{-9}-1.4397\, 10^{-8} $\\
  \textrm{Ultra low frequency ULF } & $300-3000 Hz $&$ 1.4397\, 10^{-8}-1.4397\, 10^{-7}$\\
   \textrm{ Very low frequency VLF} &$ 3-30 kHz $&$1.4397\, 10^{-7}-1.4397\, 10^{-6}$ \\
    \textrm{Low frequency LF }&$ 30-300 kHz $&$1.4397\, 10^{-6} -1.4397\, 10^{-5}$\\
 \textrm{ Medium frequency MF }& $300 KHz-3 MHz  $& $1.4397\, 10^{-5} -1.4397\, 10^{-4}$\\
 \textrm{High frequency HF} &$ 3-30 MHz $ &$1.4397\, 10^{-4}- 1.4397\, 10^{-3}$\\
  \textrm{Very high frequency VHF} &$ 30-300 MHz  $&$ 1.4397\, 10^{-3}-1.4397\, 10^{-2}$\\
 \textrm{ Ultra high frequency UHF} &$ 300 MHz-3 GHz  $&$1.4397\, 10^{-2}-1.4397\, 10^{-1}$\\
  \textrm{Super high frequency SHF} &$ 3-30 GHz $&$ 1.4397\, 10^{-1}\, -1.4397$\\
 \textrm{ Extremely high frequency EHF }&$ 30-300 GHz $&$1.4397-14.397$\\
\textrm{  Tremendously high frequency THF} &$300 GHz-3000 GHz $ &$14.397 -143.97$\\
    \hline
\end{tabular}
\caption{Critical temperatures $T_{crit}$ for typical
radio waves, with $h/k_B=4.799\,  10^{-11}$ Kelvin per second.
} \label{tablec}
\end{table}

We see from Table \ref{tablec} that serious anomalies are detected for the
P-distribution in the case of radio waves of high frequency. $P$
becomes negative, which is absurd, for rather high temperatures, where
one expects classical physics to reign. Accordingly, one concludes
that quasi-probabilities do not exhibit a sensible classical limit in
the $P-$case, contrary to what happens in both the $W$ and $Q$ ones.
\newpage
\subsection{Statistical complexity}

The statistical complexity $C$, according to Lopez-Ruiz, Mancini,
and Calvet \cite{LMC},  is a suitable product of two quantifiers,
such that $C$ becomes minimal at the extreme situations of perfect
order or total randomness. Instead of using the prescription of  \cite{LMC}, but without violating its spirit,
 we will take one of these two  quantifiers to be Fisher's measure and  the other an entropic form, since it
 is well known that the two behave in opposite manner~\cite{frieden}. Thus:

\be C=S I= \gamma\, (1-\ln \gamma), \ee that vanishes for perfect
order or total randomness. For each particular case,
we explicitly have

\ben
C_P&=&\left(e^{\beta\hbar\omega}-1\right)\left[1-\ln\left(e^{\beta\hbar\omega}-1\right)\right],\\\cr
C_Q&=&\left(1-e^{-\beta\hbar\omega}\right)\left[1-\ln\left(1-e^{-\beta\hbar\omega}\right)\right],
\\\cr C_W&=&2
\tanh(\beta\hbar\omega/2)\left[1-\ln\left(2
\tanh(\beta\hbar\omega/2)\right)\right], \een for, respectively,
the distributions $f_P$, $f_Q$, and $f_W$. The maximum of the
statistical complexity occurs when $\gamma=1$ and, the associated
temperature values are
\begin{displaymath}
 \left\{ \begin{array}{ll}
e^{\beta\hbar\omega}-1=1\Rightarrow T=\hbar\omega/k_B\ln2 >T_{crit}&\,\,\,\, \textrm{for the $f_P-$function},\\\\
1-e^{-\beta\hbar\omega} =1\Rightarrow T=0& \,\,\,\,\textrm{for the $f_Q-$function},\\\\
2\tanh(\beta\hbar\omega/2)=1 \Rightarrow
T=\hbar\omega/2k_B\arctan(1/2)&\,\,\,\, \textrm{for the
$f_W-$function}.
\end{array} \right.
\end{displaymath}

\begin{figure}[h]
\begin{center}
\includegraphics[scale=0.6,angle=0]{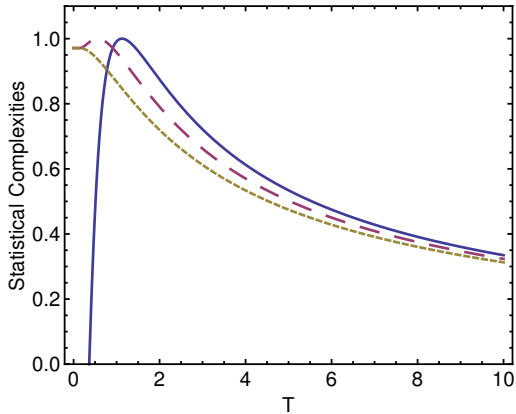}
\vspace{-0.2cm} \caption{Complexities $C_P$, $C_Q$, and $C_W$
versus the temperature $T$ in $(\hbar \omega/k_B)-$units.
The solid line is the case $P$, the dashed one the Wigner one, and the dotted curve is assigned
to the Husimi instance.}\label{complex}
\end{center}
\end{figure}

The statistical complexity $C$ is plotted in Fig. \ref{complex}.

\subsection{Linear entropy}

Another interesting information quantifier   is that of  the
Manfredi-Feix entropy \cite{manfredi}, derived  from the phase
space Tsallis $(q=2)$ entropy \cite{tsallis}. In quantum information
this form is referred to as the linear entropy \cite{sadeghi}. It
reads

\be S_l=1-\int \frac{\mathrm{d}^2 \alpha}{\pi}\,f^2(\alpha)= 1-
{\cal J}, \ee

\be  {\cal J}=\int\,\frac{\mathrm{d}^2 \alpha}{\pi}\,f^2(\alpha)=\frac{\gamma}{2}\label{jota}.
 \ee
Accordingly, we have

\be S_l=1-\frac{\gamma}{2}; \,\,\,\,\, 0 \le S_l \le 1. \ee This
is semi-classical result, valid for small $\gamma$.
In particular,

\ben
S_{l,P}&=&1-\frac{\gamma_P}{2}=1-\frac{e^{\beta\hbar\omega}-1}{2},\\\cr
S_{l,Q}&=&1-\frac{\gamma_Q}{2}=1-\frac{1-e^{-\beta\hbar\omega}}{2},\\\cr
S_{l,W}&=&1-\frac{\gamma_W}{2}=1-\frac{2
\tanh(\beta\hbar\omega/2)}{2}.
\een Note that in the $P$-instance the linear entropy becomes
negative, once again, for $T<T_{crit}$.  Contrary to what happens
for the logarithmic entropy, the linear one can vanish in the $W$
representation.

 The ensuing
statistical complexity that uses $S_l$ becomes

\be C_l= S_l\, I=I
\left(1-\frac{I}{2}\right)=\gamma\left(1-\frac{\gamma}{2}\right),
\ee
 vanishing both for $\gamma=0$ and for   $\gamma=2$, the extreme values of the $\gamma-$physical range (we showed above that $\gamma$
 cannot exceed 2 without violating uncertainty restrictions).

\begin{figure}[h]
\begin{center}
\includegraphics[scale=0.6,angle=0]{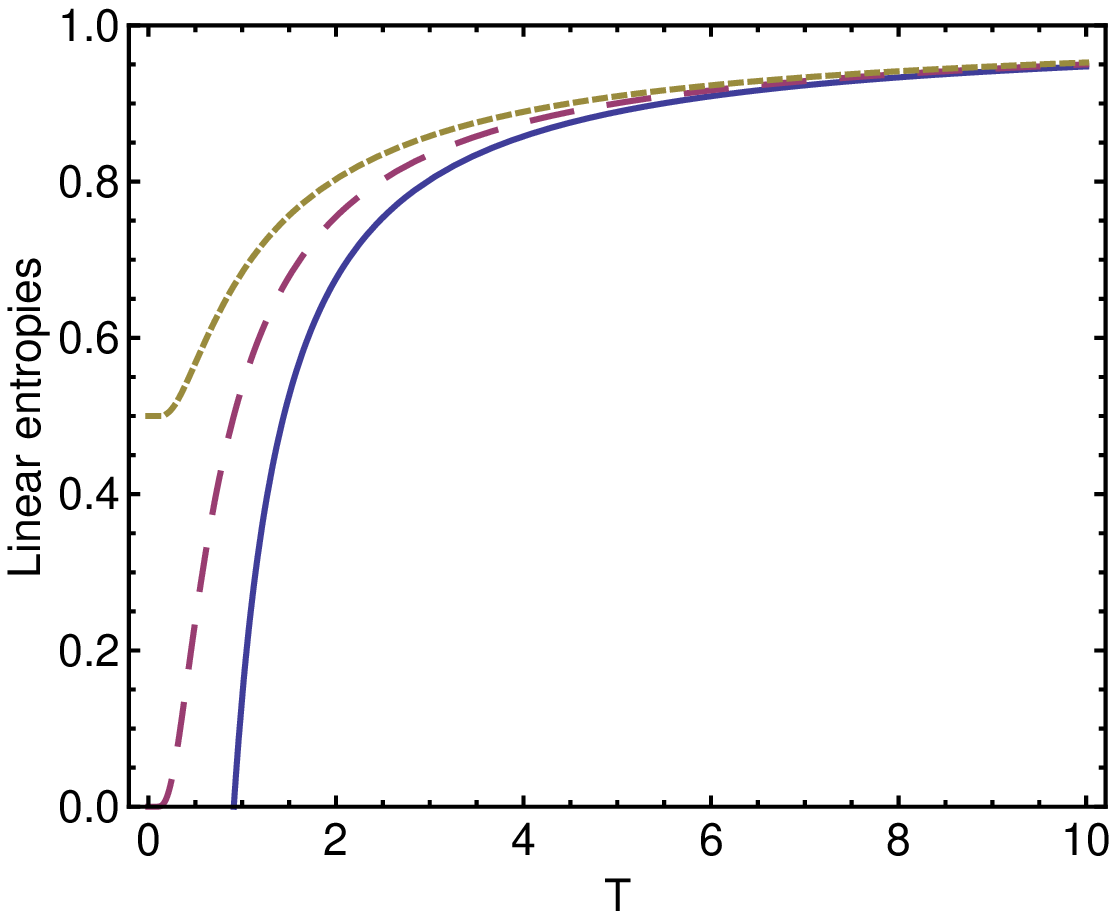} \includegraphics[scale=0.6,angle=0]{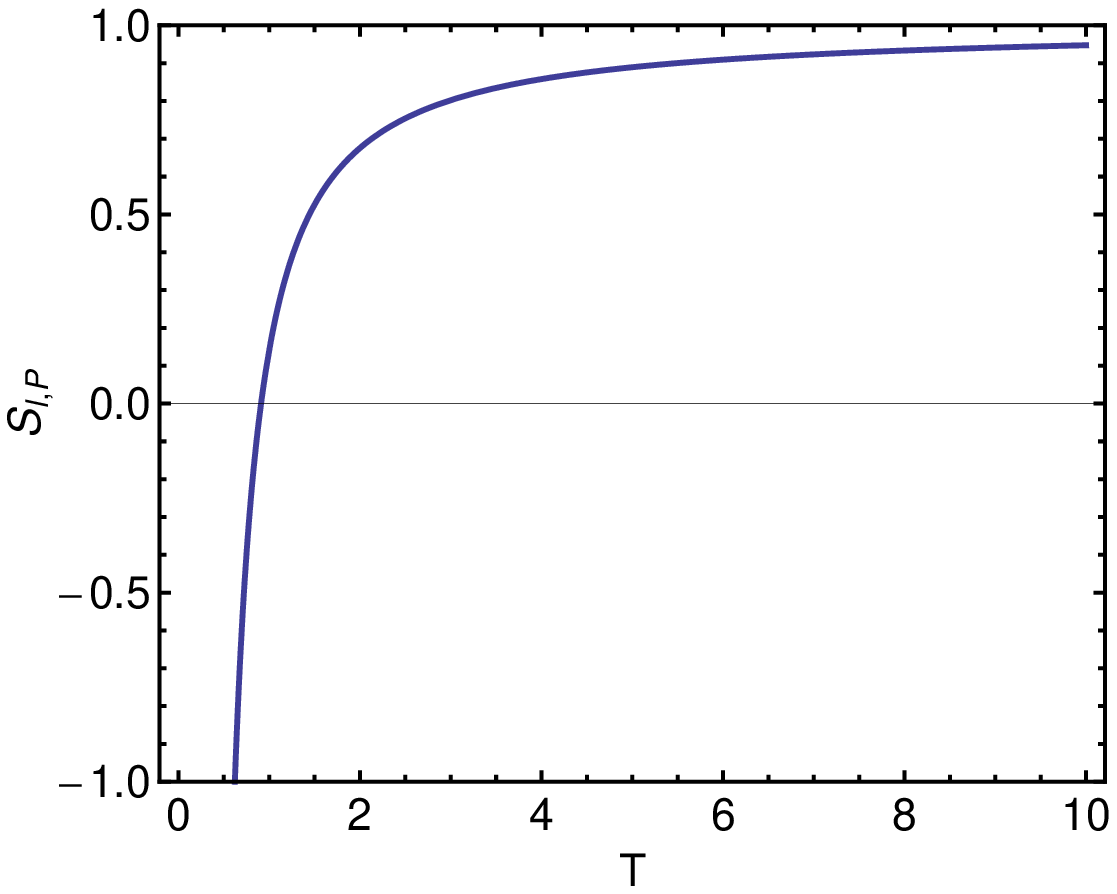}
\vspace{-0.2cm} \caption{Left: Linear entropies $S_{l,P}$,
$S_{l,Q}$, and $S_{l,W}$ versus the temperature $T$ in $(\hbar
\omega/k_B)-$units. The solid line is the case $P$, the dashed one
the Wigner one, and the dotted curve is assigned to the Husimi
instance. Right: $S_{l,P}$ as a function of the temperature $T$ in
$(\hbar \omega/k_B)-$units. We note the linear entropy $S_{l,P}$
is negative below the $T-$value $t_{crit}=T_{crit}
k_B/\hbar\omega= \ln 3$.}\label{slineal}
\end{center}
\end{figure}

\section{Fano factor's classical analog}
\label{analogfano}

In general, the Fano factor is the coefficient of
dispersion of the   probability distribution $p(y)$, which is defined as~\cite{21}

\be \mathcal{F}=\frac{\Delta y^2}{\langle y \rangle}, \ee
where $\Delta y^2=\langle y^2\rangle-\langle y\rangle^2$ is the variance and $\langle x \rangle$ is the mean of a random process $y$.

If $p(y)$ is a Poisson distribution then  one sees  that the pertinent Fano factor becomes
unity ($\mathcal{F}=1$)\cite{librazo,CBrif}.  We remind the reader
of two situations:

\begin{enumerate}
\item for $\mathcal{F} < 1$,  sub-Poissonian processes occur, while

\item for $\mathcal{F} > 1$,  corresponds to a super-Poissonian process.
\end{enumerate}

For our gaussian distribution (\ref{gaus}), if one
sets now $y=|\alpha|^2$ one has the classical Fano-analog

\be \mathcal{F}=\frac{\langle |\alpha|^4\rangle_f-\langle
|\alpha|^2\rangle_f^2}{\langle |\alpha|^2\rangle_f},\label{f1} \ee
where  the expectation
value of the function $\mathcal{A}(\alpha)$ is calculated as

\be \langle \mathcal{A}\rangle_f=\int\frac{\mathrm{d}^2
\alpha}{\pi}\,f(\alpha)\,\mathcal{A}(\alpha), \label{mvalues}\ee
indicating that $f(\alpha)$ is the statistical weight function.
Thus,
after  computing the mean values involved in \newline  Eq. (\ref{f1}) by taking into account the definition~(\ref{mvalues}), the Fano factor becomes \be
\mathcal{F}=\frac{1}{\gamma} = \frac{1}{I},\label{IFanofactor} \ee
which, for a Gaussian distribution,  links the Fano factor to the
distribution's width and to the Fisher's measure $I$.
We are speaking of processes that are of a quantum nature and can not take place
in a classical environment. Thus, with reference to the critical
temperature defined in Eq. (\ref{crit}) we have to deal with

\ben {\cal F}_P&=& \frac{1}{e^{\beta\hbar\omega}-1}\,\, (=1\,\,
\textrm{at}\,\, T=\frac{\hbar \omega}{k_B \ln{2}} > T_{crit})\,\,
\,\,\mathrm{for\,the}\,\, f_P\,\,\mathrm{function},\\\cr {\cal
F}_Q&=&\frac{1}{1-e^{-\beta\hbar\omega}} \,\,(=1
\,\,\mathrm{at}\,\, T=0) \,\, \,\,\mathrm{for\,the}\,\,
f_Q\,\,\mathrm{function},\\\cr {\cal
F}_W&=&\frac{1}{2\tanh(\beta\hbar\omega/2)}\,\, (=1
\,\,\mathrm{at}\,\, T=T_{crit})\,\, \,\,\mathrm{for\,the }\,\,
f_W\,\,\mathrm{function}. \een \label{3fanos}  The $f_Q-$case
reaches the {\it super to sub-Poissonian  transition}~ only at
$T=0$, while the other two cases reach it at finite temperatures.

%%%%%%%%%%%%%%%%%%%%%%%%%%%%%%%%%%%%%%%%%%%%%%%%%%%%%%%%%%%%%%%%%%%%%
\section{Fluctuations}
%%%%%%%%%%%%%%%%%%%%%%%%%%%%%%%%%%%%%%%%%%%%%%%%%%%%%%%%%%%%
\label{uncert}
 We start this section considering the classical Hamiltonian of the harmonic oscillator that reads

\be \mathcal{H}(x,p)= \hbar \omega |\alpha|^2, \label{H0}
  \ee where $x$ and $p$ are  phase space variables, $|\alpha|^2=x^2/4\sigma_x^2+p^2/\sigma_p^2$, and
 $\sigma_x^2=\hbar/2m\omega$ and $\sigma_p^2=\hbar
m\omega/2$~\cite{pathria}.

Using the definition of the mean value (\ref{mvalues}), from (\ref{H0}) we immediately
find~\cite{pennini1}

\be \langle x^2/2\sigma_x^2\rangle_f=\langle
p^2/2\sigma_p^2\rangle_f=\langle |\alpha|^2\rangle_f, \ee with \be
\langle |\alpha|^2\rangle_f=\gamma\int\frac{\mathrm{d}^2
\alpha}{\pi}\,e^{-\gamma|\alpha|^2}\,|\alpha|^2=\frac{1}{\gamma},
\ee where $\langle x\rangle_f=\langle p\rangle_f=\langle \alpha
\rangle_f=0$,  while $\gamma$ takes the
respective values $\gamma_P$, $\gamma_Q$, and $\gamma_W$. The
concomitant variances are $\Delta x^2=\langle x^2\rangle_f-\langle
x\rangle_f^2=2 \sigma_p^2/\gamma$, and $\Delta p^2=\langle
p^2\rangle_f-\langle p\rangle_f^2=2 \sigma_p^2/\gamma$. Hence, for
our general gaussian distribution one easily establishes that

\be {\cal U}= \Delta x  \Delta p=  \frac{\hbar}{\gamma},
\label{incer}\ee which shows that  $\gamma$ should be constrained
by  the restriction

\be \label{gorgama} \gamma \le 2, \ee if one wishes the inequality

\be \label{flafla} \Delta x  \Delta p \ge \hbar/2, \ee to hold.
\vskip 3mm

Specializing (\ref{incer}) for our three quasi-probability
distributions yields

\ben \Delta x\Delta
p&=&\frac{\hbar}{e^{\beta\hbar\omega}-1},\,\,\,\,\,\,\mathrm{for\,
\, } f_P\,\,\mathrm{function}, \\\cr \Delta x\Delta
p&=&\frac{\hbar}{1-e^{-\beta\hbar\omega}},\,\,\,\,\,\,\mathrm{for\,
\, }f_Q\,\,\mathrm{function},\\\cr \Delta x\Delta p&=&\frac{\hbar}{2
\tanh(\beta\hbar\omega/2)},\,\,\,\,\,\,\mathrm{for\, }f_W\,\,\mathrm{function}. \een The
restriction (\ref{flafla}) applied to the $P$-result entails that
it holds if

\be \label{complyT} T \ge \frac{\hbar \omega} {\ln{3} k_B}=
T_{crit} \approx 0.91023\, \frac{\hbar \omega} {k_B}.\ee
 Thus, the distribution $f_P$ seems again to
becomes un-physical at  temperatures lower than $T_{crit},$ for
which  (\ref{flafla}) is violated. From (\ref{incer}) we have
$\gamma=\hbar/{\cal U}$. Accordingly, if we insert this into
(\ref{Shannon}), the logarithmic entropy $S$ can be recast in
$\cal{U}-$terms via the relation

\begin{figure}[h]
\begin{center}
\includegraphics[scale=0.6,angle=0]{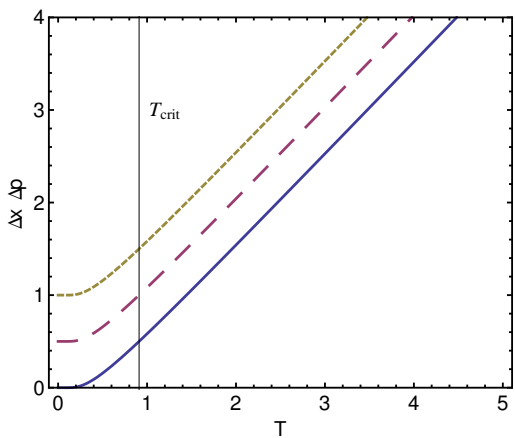}
\vspace{-0.2cm} \caption{Fluctuations vs. the temperature
$T$ in $(\hbar \omega/k_B)-$units.  The solid line is the case  $P$,
the dashed one the Wigner one, and the dotted line is assigned to
the Husimi instance. }
\end{center}
\end{figure}

\be S=1-\ln\left(\frac{\hbar}{\Delta x\Delta p}\right), \ee (also
demonstrated in Ref.~\cite{PRD2753_93} to hold for the Wehrl
entropy) that vanishes for  \be \Delta x\Delta p=\frac{\hbar}{e}.
\ee In the $P$-instance this happens at

\be T=0.71463\,\frac{\hbar\omega}{k_B}. \ee At this temperature
 the Heisenberg's-like condition
(\ref{flafla}) is violated.
 The $W$
and $Q$ distributions do not allow for such a circumstance.
Actually, in the Wigner case, which is exact, the minimum
$S-$value is attained at $\beta=\infty$, where

\be \label{MinS} S_{min}= 1-\ln{2} \approx 0.306. \ee The
uncertainty restriction (\ref{flafla}) seems to impede the
phase-space entropy to vanish,    a sort of
quasi-quantum effect. \normalcolor
 It is clear then that, in phase space, the logarithmic entropy,
by itself, is an uncertainty indicator, in agreement with the
work, in {\it other scenarios}, of several authors (see, for
instance, \cite{uff} and references therein).

\vskip 3mm

\begin{figure}[h]
\begin{center}
\includegraphics[scale=0.6,angle=0]{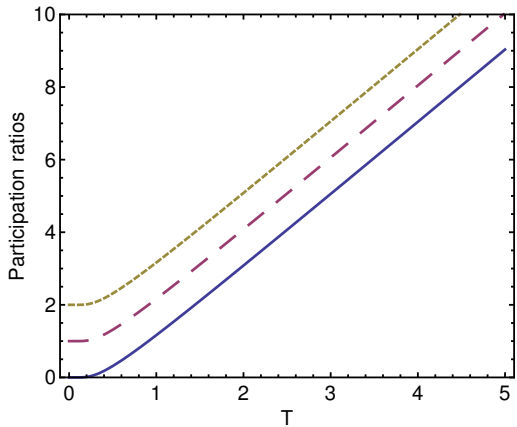}
\vspace{-0.2cm} \caption{Participation ratio $m$ versus
temperature $T$ in $(\hbar \omega/k_B)-$units. The solid line is the case $P$,
the dashed one the Wigner one, and the dotted curve is assigned
to the Husimi instance.}\label{mratio}
\end{center}
\end{figure}

Define now the participation ratio's  analog as \cite{[29],[30]}

\be m= \frac{1}{\cal J} = \frac{2}{\gamma}, \ee
 where $\mathcal{J}$ is given by (\ref{jota}). This is an
important quantity that measures the number of pure states
entering the mixture determined by our general   gaussian
probability distribution of amplitude $\gamma$  \cite{[29],[30]}.
We again encounter troubles with the $P$-distribution in this
respect. It is immediately realized by looking at
Fig.~\ref{mratio} that, for fulfilling the obvious condition $m\ge
1$, one needs a temperature $T \ge T_{crit}$.

%%\newpage
\section{Conclusions}
%%%%%%%%%%%%%%%%%%%%%%%%%%%%%%%%%%%%%%%%%%%%%%%%%%%%%%%%%%%%%%%%
\label{conclu}

We have investigated here the thermal statistics of
quasi-probabilities-analogs $f(\alpha)$ in phase space for the
important case of quadratic Hamiltonians, focusing attention on
the three more important instances, i.e., those of Wigner, $P$-,
and Husimi distributions.

\begin{itemize}

\item We emphasized the fact that for all of them the semiclassical entropy is a function
only of  the fluctuation-product $\Delta x \Delta p$. This fact allows
one to ascertain that the analog $P$-distribution seems to become
un-physical at  low enough temperatures, smaller than a critical
value $T_{crit}$, because, in such an instance:

\begin{enumerate}

\item it would  violate a Heisenberg's-like principle in such a case. The
behavior of other information quantifiers reconfirms such an
assertion, i.e.,

\item Fisher's measure exceeds its permissible maximum value
$I=2$,

\item the participation ratio becomes $< 1$, which is impossible.

\end{enumerate}

\item It is also clear then that semiclassical entropy, by itself,  in
phase space,  looks like  a kind of ``uncertainty" indicator.

\item We have determined the temperatures for which the statistical complexity becomes
maximal, as a signature of the well-known transition between
classical and nonclassical states of light whose signature is the
transition   from super-Poissonian  to sub-Poissonian
distributions \cite{zou}.

\end{itemize}

We have seen that the P-distribution in the case of radio waves of high frequency
becomes negative, which is absurd, as the pertinent temperatures are rather high and thus one expects classical physics to reign. Accordingly, one concludes
that quasi-probabilities do not exhibit a sensible classical limit in
the $P-$case, contrary to what happens in both the $W$ and $Q$ ones.

\section*{Acknowledgement}

The authors were supported by Consejo Nacional de
Investigaciones Cient\'{\i}ficas y T\'{e}cnicas (CONICET), Argentina.
 Useful discussions with Prof. R. Piasecki of Opole's
 University, Poland, are gratefully acknowledged.
%%\newpage


\begin{thebibliography}{99}

\bibitem{plato} J. Von Plato,  ``Grundbegriffe der Wahrscheinlichkeitsrechnung in Grattan-Guinness", I., ed.,
Landmark Writings in Western Mathematics, Elsevier, Amsterdam, pp 960-69, 2005.


\bibitem{sudar} E.C.G. Sudarshan,  ``Equivalence of Semiclassical and Quantum Mechanical Descriptions of Statistical Light Beams",
Physical Review Letters, vol. 10, pp. 277-279, 1963.



\bibitem{glauber} R.J. Glauber, ``Coherent and incoherent states of the radiation field", Physical Review, vol. 131, pp. 2766-2788, 1963.
%% Coherent and Incoherent States of the Radiation Field.


\bibitem{wigner} E.P. Wigner, ``On the Quantum Correction For Thermodynamic Equilibrium", Physical Review, vol. 40, pp. 749-759, 1932.
%%On the Quantum Correction For Thermodynamic Equilibrium



\bibitem{smooth} F. Pennini, A. Plastino, ``Smoothed Wigner distributions, decoherence,
and the temperature dependence of the classical-quantical frontier", The European Physical Journal D, vol. 61, pp. 241-247, 2011.
%%Smoothed Wigner distributions, decoherence, and the temperature dependence of the classical-quantical frontier


\bibitem{husimi} K. Husimi,~~``Some formal properties of the density
matrix", Proceedings of the Physico-Mathematical Society of Japan, vol. 22, pp. 264-283, 1940.
%%Some formal properties of the density matrix.
%% Pages 264-283


\bibitem{mizrahi1} S.S. Mizrahi,~~``Quantum mechanics in the Gaussian
wave-packet phase space representation", Physica A, vol. 127, pp. 241-264, 1984.

\bibitem{mizrahi2} S.S. Mizrahi,~~``Quantum mechanics in the Gaussian
wave-packet phase space representation II: Dynamics", Physica A, vol. 135, pp. 237-250, 1986.

\bibitem{mizrahi3} S.S. Mizrahi,~~``Quantum mechanics in the gaussian wave-packet phase space representation III: From phase
space probability functions to wave-functions", Physica A, vol. 150, pp. 541-554, 1988.

\bibitem{1}  Jean-Pierre Gazeau, ``Coherent states in Quantum Physics", WILEY-VCH Verlag GmbH and Co. KGaA, Weinheim, 2009.

\bibitem{lieb} E.H. Lieb, ``The classical limit of quantum spin systems", Communications in Mathematical Physics, vol. 31, pp. 327-340, 1973.

\bibitem{a6} M.H. Anderson, J.R. Ensher, M.R. Matthews, C.E. Wieman, E.A. Cornell, ``Observation of Bose-Einstein condensation in a dilute atomic vapor", Science, vol. 269, pp. 198-201, 1995 .
%% Observation of Bose-Einstein Condensation in a Dilute Atomic Vapor



\bibitem{a7} K.B. Davis, M.O. Mewes, M.R. Andrews, N.J. van Druten, D.S. Durfee, D.M. Kurn, W. Ketterle, ``Bose-Einstein condensation in a gas of sodium atoms", Physical Review Letters, vol. 75, pp. 3969-3973, 1995.
% M. -O. Mewes, M. R. Andrews, N. J. van Druten, D. S. Durfee, D. M. Kurn, and W. Ketterle
%% Bose-Einstein Condensation in a Gas of Sodium Atoms


\bibitem{a8} C.C. Bradley, C.A. Sackett, R.G. Hulet, ``Bose-Einstein condensation of lithium: Observation of limited condensate number", Physical Review Letters, vol. 78, pp. 985-989, 1997.
% CC Bradley , CA Sackett y RG Hulet
%% Bose-Einstein Condensation of Lithium: Observation of Limited Condensate Number

\bibitem{librazo} M.O. Scully, M.S. Zubairy, {\it Quantum optics}, Cambridge University Press, NY, 1997.


\bibitem{bookQOptics} W.P. Scheleich, {\it Quantum Optics in Phase Space}, Wiley VCH-Verlag, Berlin, Germany, 2001.

\bibitem{bernie} B.R. Frieden, B.H. Soffer, ``Lagrangians of physics and the Fisher transfer game", Physical Review E, vol. 52, pp. 2274-2286, 1995.

\bibitem{hall}  M.J.W. Hall, ``Quantum properties of classical Fisher information", Physical Review A, vol. 62, pp. 012107-1-012107-6, 2000.

 \bibitem{Entropy2014} F. Pennini, A. Plastino, ``Fluctuations, Entropic quantifiers and classical quantum transition", Entropy, vol. 16, pp. 1178-1190, 2014.
%%pags. 1178-1190
%%fluctuations, Entropic quantifiers and classical quantum transition

\bibitem{lieb1} E.H. Lieb, ``Proof of an Entropy Conjecture of Wehrl", Communication in Mathematical Physics, Vol. 62, pp. 35�41, (1978).

\bibitem{21}  U. Fano, "Ionization Yield of Radiations. II. The Fluctuations of the Number of Ions", {\it Physical Review} {\bf 72} 26-29, 1947.
%% "Ionization Yield of Radiations. II. The Fluctuations of the Number of Ions"
%% Pags. 26-29

\bibitem{LMC} R. L\'{o}pez-Ruiz, H.L. Mancini, X.A. Calbet, ``A statistical measure of complexity", Phys. Lett. A, vol. 209, pp. 321-326, 1995.


\bibitem{frieden} B.R. Frieden, ``Science from Fisher Information", 2 ed., Cambridge University Press, Cambridge, UK, 2008.

\bibitem{manfredi} G. Manfredi, M.R. Feix, ``Entropy and Wigner functions, Physical Review E, vol. 62, 4665-4674, 2000.


\bibitem{tsallis} C. Tsallis, ``Introduction to Nonextensive Statistical Mechanics", Springer, New York, 2008.



\bibitem{sadeghi} P. Sadeghi, S. Khademi, A.H. Darooneh, ``Tsallis entropy in phase-space quantum mechanics", Physical Review A, vol 86, pp. 012119:1--012119:8, 2012.

\bibitem{CBrif} C. Brif, Y. Ben-Aryeh, ``Subcoherent $P$-representation for non-classical photon states", Quantum Optics, vol. 6, pp. 391-396, 1994.




\bibitem{pathria} R.K. Pathria, Statistical Mechanics, Pergamon Press, Exeter, 1993.


\bibitem{pennini1} F. Pennini, A. Plastino, ``Heisenberg--Fisher
thermal uncertainty measure",  Physical Review~E, vol. 69, pp. 057101:1-057101:4, 2004.
%% Heisenberg--Fisher thermal uncertainty measure.
%% 4 pages

\bibitem{PRD2753_93} A. Anderson, J.J. Halliwell, ``Information-theoretic measure of uncertainty due to quantum and thermal fluctuations", Physical~Review~D, vol48, pp. 2753-2765, 1993.
%% Information-theoretic measure of uncertainty due to quantum and thermal fluctuations.



\bibitem{uff}  I. Bialynicki-Birula, J. Mycielski, ``Uncertainty Relations for Information Entropy in Wave Mechanics", Communications in Mathematical Physics, vol. 44, pp. 129-132, 1975.


\bibitem{[29]} J. Batle, A.R. Plastino, M. Casas, A. Plastino, ``Conditional $q$-entropies and quantum separability: a numerical exploration", Journal of Physics A: Mathematical and General, vol. 35, pp. 10311-10324, 2002.

\bibitem{[30]} F. Pennini, A. Plastino, G.L. Ferri, ``Semiclassical information from deformed and escort information measures", Physica A, vol. 383, pp. 782-796, 2007.


\bibitem{zou}  X.T. Zou, L. Mandel, ``Photon-antibunching and sub-Poissonian photon statistics",  Physical  Review A, vol. 41, pp. 475-476,~1990.



\end{thebibliography}
\end{document}